\documentclass[conference]{IEEEtran}
\IEEEoverridecommandlockouts

\usepackage{cite}
\usepackage{amsmath,amssymb,amsfonts}
\usepackage{algorithmic}
\usepackage{graphicx}
\usepackage{textcomp}
\usepackage{xcolor}
\usepackage{array}
\usepackage{verbatim}
\usepackage{multirow}
\usepackage{amsbsy}
\usepackage{amstext}
\def\BibTeX{{\rm B\kern-.05em{\sc i\kern-.025em b}\kern-.08em
    T\kern-.1667em\lower.7ex\hbox{E}\kern-.125emX}}
\begin{document}

\title{Modified Delayed Acceptance MCMC for Quasi-Bayesian Inference with Linear Moment Conditions
\thanks{This work was supported by JSPS KAKENHI Grant Number 25K21168.}
}

\author{\IEEEauthorblockN{Masahiro Tanaka}
\IEEEauthorblockA{\textit{Faculty of Economics} \\
\textit{Fukuoka University}\\
Fukuoka, Japan \\
gspddlnit45@toki.waseda.jp}
}

\maketitle

\begin{abstract}
We develop a computationally efficient framework for quasi-Bayesian
inference based on linear moment conditions. The approach employs
a delayed acceptance Markov chain Monte Carlo (DA--MCMC) algorithm
that uses a surrogate target kernel and a proposal distribution derived
from an approximate conditional posterior, thereby exploiting the
structure of the quasi-likelihood. Two implementations are introduced.
DA--MCMC--Exact fully incorporates prior information into the proposal
distribution and maximizes per-iteration efficiency. Conversely, DA--MCMC--Approx
omits the prior in the proposal to reduce matrix inversions, thus
improving numerical stability and computational speed in higher-dimensional
settings. Simulation studies on heteroskedastic linear regression
models demonstrate substantial gains over both standard MCMC and conventional
DA--MCMC methods, as measured by multivariate effective sample size
per iteration and per second. The Approx variant delivers the highest
overall throughput, whereas the Exact variant attains the highest
per-iteration efficiency. Applications to two empirical instrumental
variable regressions corroborate these findings: the Approx implementation
scales effectively to larger designs where other methods become impractical,
while still delivering precise inference. Although developed for moment-based
quasi-posteriors, the proposed approach also extends naturally to
risk-based quasi-Bayesian formulations when first-order conditions
are linear and support analogous transformations. Overall, the framework
offers a practical, scalable, and robust tool for conducting quasi-Bayesian
analysis across a wide range of statistical applications.
\end{abstract}


\section{Introduction}

Bayesian analysis offers a coherent and flexible framework for inference,
enabling principled uncertainty quantification through a combination
of prior information and data. However, conventional Bayesian methods
require full specification of the underlying probabilistic model,
making Bayesian inference vulnerable to model misspecification. To
mitigate this issue, a growing body of research has explored quasi-Bayesian
approaches that relax the requirement for an exact likelihood specification.
These approaches construct alternative quasi-likelihoods based on
loss functions \cite{Zhang2006a,zhang2006b,Jiang2008,Bissiri2016,Syring2023}
or moment conditions \cite{Kim2002,Chernozhukov2003,Yin2009}, thereby
enhancing robustness while preserving the interpretability of Bayesian
posterior inference.

The present study adopts a quasi-likelihood formulation derived from
the generalized method of moments (GMM) criterion \cite{Hansen1982,Hall2004}.
Within the GMM framework, the statistical model is defined through
a set of moment conditions rather than an explicit likelihood function.
This formulation allows inference to proceed without stringent distributional
assumptions, such as error normality or functional form restrictions,
while still enabling probabilistic interpretation through the quasi-posterior.
Posterior inference can then be carried out using simulation-based
techniques such as the Markov chain Monte Carlo (MCMC) method \cite{Kim2002,Chernozhukov2003,Yin2009,Hong2021}.

However, quasi-Bayesian inference based on moment conditions presents
substantial computational challenges. Evaluating the quasi-posterior
typically involves repeated matrix inversions and determinant calculations,
which are both computationally intensive and prone to numerical instability---especially
in high-dimensional settings. Existing work, such as the sampler proposed
by \cite{Yin2011}, sought to improve numerical stability but often
at significant computational cost. More recently, \cite{Tanakafothcoming}
introduced a delayed acceptance MCMC (DA--MCMC) algorithm \cite{Christen2005}
specifically designed for moment-based quasi-Bayesian inference. This
approach demonstrated clear efficiency gains over conventional quasi-posterior
simulation methods, establishing DA--MCMC as a promising direction
for computation in moment-based quasi-Bayesian inference. Nonetheless,
the improvement remains modest, and the fundamental computational
difficulty persists, particularly when dealing with high-dimensional
or weakly identified models.

To address these limitations, this paper builds on prior work and
proposes an efficient framework for quasi-Bayesian inference based
on the GMM criterion. The framework uses a modified DA--MCMC algorithm
tailored to quasi-posteriors derived from linear moment conditions.
The proposed method leverages the linear structure of the moment equations
to design a computationally tractable proposal distribution that closely
approximates the target posterior. Two implementations of the algorithm
are introduced, balancing the trade-off between computational efficiency
and numerical stability. Through simulation experiments and empirical
applications, we demonstrate that the proposed framework achieves
substantial improvements in both sampling efficiency and computational
speed.

More broadly, the proposed framework also relates to recent quasi-Bayesian
approaches based on empirical risk functions, such as that of \cite{Frazier2024}.
Although their quasi-posterior formulation arises from a general decision-theoretic
framework rather than explicit moment conditions, the method proposed
in this paper can be applied to such models whenever the first-order
condition of the empirical risk function is linear in the parameters
and can be transformed analogously to the moment condition structure
considered here. This connection highlights the broader applicability
of the proposed algorithm beyond the GMM-based context.

The remainder of the paper is organized as follows. Section II introduces
the quasi-Bayesian framework based on moment conditions and develops
the proposed DA--MCMC algorithms. Section III presents simulation
studies using synthetic data to evaluate the computational performance
of the proposed methods. Section IV applies the framework to real-world
datasets to demonstrate its empirical relevance. Section V concludes
with a summary of the findings and discusses directions for future
research. For quick reference, Table I summarizes the key symbols
and hyperparameters used throughout the paper.

\begin{table}

\caption{Key symbols and hyperparameters used in the study}

\medskip{}

\begin{centering}
\begin{tabular}{cl}
\hline 
Symbol & Description\tabularnewline
\hline 
$n$ & Sample size\tabularnewline
$k$ & Number of unknown parameters\tabularnewline
$\mathcal{D}_{i}$ & Set of observations for the $i$th instance\tabularnewline
$\boldsymbol{\theta}$ & Vector of unknown parameters\tabularnewline
$\boldsymbol{m}_{i}\left(\cdot,\cdot\right)$ & Moment function\tabularnewline
$\bar{\boldsymbol{m}}\left(\cdot,\cdot\right)$ & Empirical mean of $\boldsymbol{m}_{i}\left(\cdot,\cdot\right)$\tabularnewline
$\pi\left(\cdot\right)$ & Quasi-posterior kernel\tabularnewline
$\pi^{*}\left(\cdot\right)$ & Surrogate kernel of $\pi\left(\cdot\right)$\tabularnewline
$\boldsymbol{W}$ & Weighting matrix\tabularnewline
$q_{1}\left(\cdot|\cdot\right)$ & Proposal distribution\tabularnewline
$\alpha_{j}\left(\cdot,\cdot\right)$ & $j$th-stage acceptance probability\tabularnewline
$\boldsymbol{\theta}^{\dagger}$,$\boldsymbol{G}$ & Task-specific quantities (see Sections II.C and IV)\tabularnewline
$\boldsymbol{Q}$ & Matrix related to the prior precision\tabularnewline
$\boldsymbol{\tau}$ & Set of hyperparameters\tabularnewline
\hline 
\end{tabular}
\par\end{centering}
\end{table}

\section{Method}

\subsection{Quasi-posterior}

A statistical model is inferred from the moment condition $\mathbb{E}\left[\boldsymbol{m}_{i}\left(\boldsymbol{\theta},\mathcal{D}_{i}\right)\right]=\boldsymbol{0}_{k}$,
where $\boldsymbol{\theta}$ denotes a $k$-dimensional vector of
unknown parameters, $\mathcal{D}_{i}$ denotes a set of observations,
$\boldsymbol{m}_{i}\left(\cdot\right)$ is a vector-valued function
referred to as the moment function, and $\boldsymbol{0}_{a}$ denotes
an $a$-dimensional zero vector. We assume that $\boldsymbol{m}_{i}\left(\cdot,\cdot\right)$
has dimension $k$---that is, the model is exactly identified---and
that $\boldsymbol{m}_{i}\left(\cdot,\cdot\right)$ is linear in $\boldsymbol{\theta}$.
Specifically, this paper focuses on moment functions of the form
\[
\boldsymbol{m}_{i}\left(\boldsymbol{\theta},\mathcal{D}_{i}\right)=\boldsymbol{A}\left(\mathcal{D}_{i}\right)\boldsymbol{\theta}-\boldsymbol{b}\left(\mathcal{D}_{i}\right)
\]
where $\boldsymbol{A}\left(\cdot\right)$ and $\boldsymbol{b}\left(\cdot\right)$
are known functions. The assumption of exact identification guarantees
the existence of a unique solution to the sample moment equations
\[
\frac{1}{n}\sum_{i=1}^{n}\boldsymbol{m}_{i}\left(\boldsymbol{\theta},\mathcal{D}_{i}\right)=\boldsymbol{0}_{k}.
\]
For brevity, we omit the dependence of $\boldsymbol{m}_{i}\left(\boldsymbol{\theta},\mathcal{D}_{i}\right)$
on $\mathcal{D}_{i}$ and simply write $\boldsymbol{m}_{i}\left(\boldsymbol{\theta}\right)$
in what follows.

The proposed framework encompasses a wide range of statistical models.
For instance, a standard linear regression can be written as 
\begin{equation}
y_{i}=\boldsymbol{\theta}^{\top}\boldsymbol{x}_{i}+u_{i},\label{eq:=000020linear=000020regression}
\end{equation}
where $y_{i}$ denotes an outcome variable, $\boldsymbol{x}_{i}$
is a $k$-dimensional vector of covariates, $\boldsymbol{\theta}$
represents the corresponding coefficients, and $u_{i}$ denotes an
error term. The model is estimated based on the following moment condition:
\[
\mathbb{E}\left[\left(y_{i}-\boldsymbol{\theta}^{\top}\boldsymbol{x}_{i}\right)\boldsymbol{x}_{i}\right]=\boldsymbol{0}_{k}.
\]
Under this approach, the distribution of the error terms is not assumed,
making it more robust to model misspecification than the standard
Bayesian approach. A linear probability model for a binary outcome
can be formulated analogously. Multivariate regression models, such
as seemingly unrelated regression \cite{Zellner1962,Fiebig2001} and
the local projection model \cite{Jorda2005,Jorda2025}, can also be
treated in a similar manner\footnote{Although a local projection model is designed for time series data,
it can be regarded as a model for independent observations as long
as a sufficient number of lagged responses are included \cite{MontielOlea2021}.}. 

Another important class of models is the instrumental variable (IV)
regression model \cite{Angrist2001,Burgess2017}, which consists of
two equations: 
\begin{eqnarray}
x_{i}^{\textrm{treat}} & = & g\left(x_{i}^{\textrm{instr}},\tilde{\boldsymbol{x}}_{i}\right)+v_{i},\nonumber \\
y_{i} & = & \theta^{\textrm{treat}}x_{i}^{\textrm{treat}}+\tilde{\boldsymbol{\theta}}^{\top}\tilde{\boldsymbol{x}}_{i}+u_{i},\label{eq:=000020IV=000020regression}
\end{eqnarray}
where $y_{i}$ denotes an outcome variable, $x_{i}^{\textrm{treat}}$
represents a treatment variable, $x_{i}^{\textrm{instr}}$ denotes
an instrumental variable that is correlated with $x_{i}^{\textrm{treat}}$
but affects $y_{i}$ only through $x_{i}^{\textrm{treat}}$, $\tilde{\boldsymbol{x}}_{i}$
is a vector of covariates, $g\left(\cdot\right)$ denotes an unknown
function, $\theta^{\textrm{treat}}$ and $\tilde{\boldsymbol{\theta}}$
are unknown parameters, and $v_{i}$ and $u_{i}$ are error terms.
Define 
\[
\boldsymbol{x}_{i}=\left(\begin{array}{c}
x_{i}^{\textrm{treat}}\\
\tilde{\boldsymbol{x}}_{i}
\end{array}\right),\quad\boldsymbol{z}_{i}=\left(\begin{array}{c}
x_{i}^{\textrm{instr}}\\
\tilde{\boldsymbol{x}}_{i}
\end{array}\right),\quad\boldsymbol{\theta}=\left(\begin{array}{c}
\theta^{\textrm{treat}}\\
\tilde{\boldsymbol{\theta}}
\end{array}\right).
\]
Then, the model can be estimated using the moment condition
\[
\mathbb{E}\left[\left(y_{i}-\boldsymbol{\theta}^{\top}\boldsymbol{x}_{i}\right)\boldsymbol{z}_{i}\right]=\boldsymbol{0}_{k}.
\]
This approach offers several advantages over conventional Bayesian
instrumental variable regression methods. It eliminates the need to
estimate the first-stage regression function and imposes no assumptions
on the distributional form of the error terms. Some measurement error
(error-in-variables) models \cite{Fuller2009} can also be formulated
in a similar manner.

We construct the quasi-likelihood based on the generalized method
of moments criterion \cite{Hansen1982,Hall2004}. Given a prior distribution
$p\left(\boldsymbol{\theta}\right)$, the quasi-posterior kernel---that
is, the quasi-posterior density evaluated up to the normalizing constant---is
specified as
\[
\pi\left(\boldsymbol{\theta}\right)=\left|\boldsymbol{W}\right|^{\frac{1}{2}}\exp\left(-\frac{n}{2}\bar{\boldsymbol{m}}\left(\boldsymbol{\theta}\right)^{\top}\boldsymbol{W}\bar{\boldsymbol{m}}\left(\boldsymbol{\theta}\right)\right)p\left(\boldsymbol{\theta}\right),
\]
where
\[
\bar{\boldsymbol{m}}\left(\boldsymbol{\theta}\right)=\frac{1}{n}\sum_{i=1}^{n}\boldsymbol{m}_{i}\left(\boldsymbol{\theta}\right),
\]
denotes the empirical mean of the moment function $\boldsymbol{m}_{i}\left(\boldsymbol{\theta}\right)$,
$n$ denotes the sample size, and $\boldsymbol{W}$ is a symmetric
positive-definite weighting matrix. 

Under certain conditions, our framework is closely related to the
approach of \cite{Frazier2024}, where an inferential problem is formulated
based on a loss function $r\left(\cdot,\cdot\right)$ as 
\[
\min_{\boldsymbol{\theta}}\sum_{i=1}^{n}r\left(\boldsymbol{\theta},\mathcal{D}_{i}\right).
\]
In that setting, \cite{Frazier2024} constructed the quasi-posterior
by defining the moment function based on the first-order condition
of the loss function,
\[
\boldsymbol{m}_{i}\left(\boldsymbol{\theta}\right)=\nabla_{\boldsymbol{\theta}}r\left(\boldsymbol{\theta},\mathcal{D}_{i}\right).
\]
The gradient of their empirical risk function thus plays a role analogous
to the empirical mean of the moment function in our framework. Consequently,
our methods apply whenever the first-order condition of the loss function
is linear in the parameters, such as in linear regression with a quadratic
loss.

The weighting matrix $\boldsymbol{W}$ is specified as the inverse
of the empirical covariance matrix of $\boldsymbol{m}_{i}\left(\boldsymbol{\theta}\right)$:
\[
\boldsymbol{W}=\boldsymbol{V}^{-1},
\]
\[
\boldsymbol{V}=\frac{1}{n-1}\sum_{i=1}^{n}\left(\boldsymbol{m}_{i}\left(\boldsymbol{\theta}\right)-\bar{\boldsymbol{m}}\left(\boldsymbol{\theta}\right)\right)\left(\boldsymbol{m}_{i}\left(\boldsymbol{\theta}\right)-\bar{\boldsymbol{m}}\left(\boldsymbol{\theta}\right)\right)^{\top}.
\]
This choice is appealing because the resulting estimator produces
the smallest credible set, which attains its nominal asymptotic coverage.
For instance, a 90\% credible interval is expected to contain the
true parameter with probability approaching 90\% as the sample size
tends to infinity \cite{Frazier2024}. 

The main obstacle to implementing this inferential strategy is the
associated computational cost and the potential for numerical instability.
When the weighting matrix $\boldsymbol{W}$ is treated as a function
of $\boldsymbol{\theta}$, the quasi-posterior is expressed as
\begin{eqnarray}
\pi\left(\boldsymbol{\theta}\right) & = & \left|\boldsymbol{W}\left(\boldsymbol{\theta}\right)\right|^{\frac{1}{2}}\times\nonumber \\
 &  & \exp\left(-\frac{n}{2}\bar{\boldsymbol{m}}\left(\boldsymbol{\theta}\right)^{\top}\boldsymbol{W}\left(\boldsymbol{\theta}\right)\bar{\boldsymbol{m}}\left(\boldsymbol{\theta}\right)\right)p\left(\boldsymbol{\theta}\right).\label{eq:=000020quasi-posterior}
\end{eqnarray}
Posterior simulation from this target kernel is known to be computationally
inefficient and numerically unstable \cite{Tanaka2021}. 

\subsection{DA--MCMC}

We propose an algorithm based on the DA--MCMC method \cite{Christen2005}.
The DA--MCMC method is a variant of the Metropolis--Hastings algorithm
that incorporates a screening process. In the first stage, given the
current state $\boldsymbol{\theta}^{\left(t\right)}$, a proposal
for the new state $\boldsymbol{\theta}^{\prime}$ is drawn from a
proposal distribution $q_{1}\left(\cdot|\boldsymbol{\theta}^{\left(t\right)}\right)$.
The proposal $\boldsymbol{\theta}^{\prime}$ is then evaluated based
on a surrogate kernel, denoted by $\pi^{*}\left(\cdot\right)$, which
serves as a computationally inexpensive approximation to the target
kernel $\pi\left(\cdot\right)$. The proposal is accepted with probability
\[
\alpha_{1}\left(\boldsymbol{\theta}^{\left(t\right)},\boldsymbol{\theta}^{\prime}\right)=\min\left\{ 1,\;\frac{q_{1}\left(\boldsymbol{\theta}^{\left(t\right)}|\boldsymbol{\theta}^{\prime}\right)\pi^{*}\left(\boldsymbol{\theta}^{\prime}\right)}{q_{1}\left(\boldsymbol{\theta}^{\prime}|\boldsymbol{\theta}^{\left(t\right)}\right)\pi^{*}\left(\boldsymbol{\theta}^{\left(t\right)}\right)}\right\} .
\]
If $\boldsymbol{\theta}^{\prime}$ is rejected, the current state
is retained, i.e., $\boldsymbol{\theta}^{\left(t+1\right)}=\boldsymbol{\theta}^{\left(t\right)}$.
If accepted, $\boldsymbol{\theta}^{\prime}$ advances to the second
stage, where it is accepted with probability
\[
\alpha_{2}\left(\boldsymbol{\theta}^{\left(t\right)},\boldsymbol{\theta}^{\prime}\right)=\min\left\{ 1,\;\frac{q_{2}\left(\boldsymbol{\theta}^{\left(t\right)}|\boldsymbol{\theta}^{\prime}\right)\pi\left(\boldsymbol{\theta}^{\prime}\right)}{q_{2}\left(\boldsymbol{\theta}^{\prime}|\boldsymbol{\theta}^{\left(t\right)}\right)\pi\left(\boldsymbol{\theta}^{\left(t\right)}\right)}\right\} .
\]
Here, the effective second-stage proposal distribution is defined
as
\[
q_{2}\left(\boldsymbol{\theta}^{\prime}|\boldsymbol{\theta}^{\left(t\right)}\right)=\alpha_{1}\left(\boldsymbol{\theta}^{\left(t\right)},\boldsymbol{\theta}^{\prime}\right)q_{1}\left(\boldsymbol{\theta}^{\prime}|\boldsymbol{\theta}^{\left(t\right)}\right).
\]
In particular, \cite{Tanakafothcoming} specifies the surrogate kernel
by replacing $\boldsymbol{W}\left(\boldsymbol{\theta}^{\prime}\right)$
with $\boldsymbol{W}\left(\boldsymbol{\theta}^{\left(t\right)}\right)$,
yielding
\begin{eqnarray*}
\pi^{*}\left(\boldsymbol{\theta}^{\prime}\right) & = & \left|\boldsymbol{W}\left(\boldsymbol{\theta}^{\left(t\right)}\right)\right|^{\frac{1}{2}}\times\\
 &  & \exp\left(-\frac{n}{2}\bar{\boldsymbol{m}}\left(\boldsymbol{\theta}^{\prime}\right)^{\top}\boldsymbol{W}\left(\boldsymbol{\theta}^{\left(t\right)}\right)\bar{\boldsymbol{m}}\left(\boldsymbol{\theta}^{\prime}\right)\right)p\left(\boldsymbol{\theta}^{\prime}\right).
\end{eqnarray*}
As noted by \cite{Christen2005}, each iteration of the DA--MCMC
method is typically slightly less efficient than the standard MCMC
when efficiency is measured by effective sample size per iteration.
However, the DA--MCMC method can achieve higher overall efficiency
in terms of effective sample size per unit time, as it avoids the
principal bottleneck: repeated evaluations of $\boldsymbol{W}$ and
$\left|\boldsymbol{W}\right|$. 

Previous studies \cite{Christen2005,Tanakafothcoming} use a multivariate
normal proposal distribution that is independent of the target kernel,
which makes the DA--MCMC algorithm a close variant of the random-walk
Metropolis--Hastings algorithm. Although this choice ensures broad
applicability, it is computationally inefficient. 

\subsection{Modified DA--MCMC}

To address this limitation, the present study replaces the generic
multivariate normal proposal with an approximate conditional posterior
distribution that leverages the linear structure of the target kernel.
To illustrate the concept, we focus on the linear regression model
in (\ref{eq:=000020linear=000020regression}). We define 
\[
\boldsymbol{y}=\left(y_{1},...,y_{n}\right)^{\top},\quad\boldsymbol{X}=\left(\boldsymbol{x}_{1},...,\boldsymbol{x}_{n}\right)^{\top}.
\]
 Then, the core component of the quasi-posterior (\ref{eq:=000020quasi-posterior})
can be expressed as
\[
\exp\left(-\frac{n}{2}\bar{\boldsymbol{m}}\left(\boldsymbol{\theta}\right)^{\top}\boldsymbol{W}\left(\boldsymbol{\theta}\right)\bar{\boldsymbol{m}}\left(\boldsymbol{\theta}\right)\right)
\]
\[
=\exp\left(-\frac{n}{2}\left[\frac{1}{n}\boldsymbol{X}^{\top}\left(\boldsymbol{y}-\boldsymbol{X}\boldsymbol{\theta}\right)\right]^{\top}\boldsymbol{W}\left[\frac{1}{n}\boldsymbol{X}^{\top}\left(\boldsymbol{y}-\boldsymbol{X}\boldsymbol{\theta}\right)\right]\right)
\]
\begin{equation}
\propto\exp\left(-\frac{n}{2}\left(\boldsymbol{\theta}-\hat{\boldsymbol{\theta}}^{\dagger}\right)^{\top}\boldsymbol{G}^{\top}\boldsymbol{W}\boldsymbol{G}\left(\boldsymbol{\theta}-\hat{\boldsymbol{\theta}}^{\dagger}\right)\right),\label{eq:=000020transformed=000020quasi-posterior}
\end{equation}
where $\hat{\boldsymbol{\theta}}^{\dagger}=\left(\boldsymbol{X}^{\top}\boldsymbol{X}\right)^{-1}\boldsymbol{X}^{\top}\boldsymbol{y}$
denotes the ordinary least squares estimator for $\boldsymbol{\theta}$
and $\boldsymbol{G}=n^{-1}\boldsymbol{X}^{\top}\boldsymbol{X}$ (See
the Appendix for the derivation). Assume that the (conditional) prior
density is specified as
\[
p\left(\boldsymbol{\theta}\right)\propto\exp\left(-\frac{1}{2}\boldsymbol{\theta}^{\top}\boldsymbol{Q}\boldsymbol{\theta}\right),
\]
where $\boldsymbol{Q}$ is a symmetric matrix parameterized by a set
of hyperparameters $\boldsymbol{\tau}$. The (conditional) posterior
density of $\boldsymbol{\theta}$ is then expressed as $N\left(\boldsymbol{\theta}|\boldsymbol{\Omega}\boldsymbol{\Upsilon}\hat{\boldsymbol{\theta}}^{\dagger},\boldsymbol{\Omega}\right),$
where $\boldsymbol{\Omega}=\left(\boldsymbol{\Upsilon}+\boldsymbol{Q}\right)^{-1}$,
$\boldsymbol{\Upsilon}=n\boldsymbol{G}^{\top}\boldsymbol{W}\boldsymbol{G}$,
and $N\left(\boldsymbol{a}|\boldsymbol{b},\boldsymbol{C}\right)$
represents the probability density function of a multivariate normal
distribution with mean $\boldsymbol{b}$ and covariance matrix $\boldsymbol{C}$,
evaluated at $\boldsymbol{a}$. Using this (conditional) posterior
distribution, we specify the first-stage proposal distribution as
\[
q_{1}\left(\boldsymbol{\theta}^{\prime}|\boldsymbol{\theta}^{\left(t\right)}\right)=N\left(\boldsymbol{\theta}^{\prime}|\boldsymbol{\Omega}\boldsymbol{\Upsilon}\hat{\boldsymbol{\theta}}^{\dagger},\boldsymbol{\Omega}\right).
\]

The direct implementation, hereafter referred to as DA--MCMC--Exact,
requires repeated matrix inversions, $\boldsymbol{\Omega}=\left(\boldsymbol{\Upsilon}+\boldsymbol{Q}\right)^{-1}$.
The term \textit{Exact} highlights that this version fully incorporates
prior information into the proposal distribution. By doing so, this
approach substantially enhances the efficiency of posterior simulations.

However, this version incurs both high computational cost and potential
numerical instability. Ensuring that $\boldsymbol{\Omega}$ remains
invertible across the sampling space of $\left(\boldsymbol{\theta},\boldsymbol{\tau}\right)$
is often difficult. In some cases, $\boldsymbol{Q}$ acts as a regularizer
for $\boldsymbol{\Upsilon}$, as in a Tikhonov inverse, which stabilizes
the simulation. In other cases, however, variability in $\boldsymbol{Q}$
introduces numerical instability---particularly when employing shrinkage
priors \cite{Park2008,Carvalho2010,Bhattacharya2015} that induce
substantial fluctuations in $\boldsymbol{\tau}$.

To address this problem, we introduce an alternative implementation.
This variant excludes prior information from the proposal distribution
and sets $\boldsymbol{\Omega}=\boldsymbol{\Upsilon}^{-1}$, which
leads to
\[
q_{1}\left(\boldsymbol{\theta}^{\prime}|\boldsymbol{\theta}^{\left(t\right)}\right)=N\left(\boldsymbol{\theta}^{\prime}|\hat{\boldsymbol{\theta}}^{\dagger},\boldsymbol{\Upsilon}^{-1}\right).
\]
This alternative implementation is hereafter referred to as DA--MCMC--Approx.
The term \textit{Approx} indicates that this version approximates
the conditional posterior by omitting prior information from the proposal
distribution. 

The relative performance of DA--MCMC--Exact and DA--MCMC--Approx
depends on the context, specifically the balance between the quasi-likelihood
and the prior in shaping the quasi-posterior. The Approx version performs
well when the quasi-likelihood dominates the quasi-posterior or when
the prior is sufficiently non-informative. In contrast, its performance
may deteriorate when the quasi-likelihood provides limited information,
such as in small-sample settings, or when the prior exerts a strong
influence on the quasi-posterior.

\section{Application to Synthetic Data}

We applied the proposed approach to infer a heteroskedastic linear
regression model using synthetic data under various scenarios. This
application compared the two implementations of the proposed approach
with two established benchmark methods. The first benchmark was the
adaptive random-walk Metropolis--Hastings algorithm, specifically
the version of \cite{Vihola2012}. The second benchmark was the DA--MCMC
algorithm of \cite{Tanakafothcoming}, in which the proposal distribution
is multivariate normal with an adaptively chosen covariance matrix.
For both benchmark methods, the tuning parameters---namely target
acceptance rate and learning rate---were set to the same values as
those employed by \cite{Vihola2012}.

The synthetic data were generated following a design inspired by \cite{Frazier2024}.
Observations were generated from a normal distribution with non-constant
variance, $y_{i}\sim\mathcal{N}\left(\boldsymbol{\theta}^{\top}\boldsymbol{x}_{i},\sigma_{i}^{2}\right)$.
Each covariate vector $\boldsymbol{x}_{i}$ consisted of a constant
term and exogenous random variables, 
\[
\boldsymbol{x}_{i}=\left(1,\tilde{\boldsymbol{x}}_{i}^{\top}\right)^{\top},\quad\tilde{\boldsymbol{x}}_{i}\sim\mathcal{N}\left(\boldsymbol{0}_{k-1},\boldsymbol{S}\right).
\]
The covariance matrix $\boldsymbol{S}$ was constructed as follows.
First, a symmetric positive definite matrix was drawn from an inverse
Wishart distribution with identity scale matrix and $k+1$ degrees
of freedom, $\boldsymbol{S}\sim\mathcal{IW}\left(\boldsymbol{I}_{k-1},k+1\right)$.
It was normalized to obtain a correlation matrix: 
\[
\boldsymbol{S}\leftarrow\tilde{\boldsymbol{S}}\boldsymbol{S}\tilde{\boldsymbol{S}},\quad\tilde{\boldsymbol{S}}=\textrm{diag}\left(s_{1,1}^{-0.5},....,s_{k-1,k-1}^{-0.5}\right),
\]
 where $s_{j,j}$ denotes the $j$th diagonal entry of $\boldsymbol{S}$.
The coefficient vector was specified as $\boldsymbol{\theta}=\left(1,1,1,0,...,0\right)^{\top}$.
The variance of the error terms depended on a subset of covariates,
defined as 
\[
\sigma_{i}^{2}=\left(1+x_{i,2}^{2}+x_{i,3}^{2}\right)/3.
\]

Three prior specifications were examined. The first specification,
referred to as the Normal prior, assumes a normal prior with a constant
unit variance, $\boldsymbol{\theta}\sim\mathcal{N}\left(\boldsymbol{0}_{k},\boldsymbol{I}_{k}\right)$.
The second specification, termed the NIG-homo prior, adopts a normal-inverse-gamma
prior with a single common hyperparameter: 
\[
\boldsymbol{\theta}|\tau\sim\mathcal{N}\left(\boldsymbol{0}_{k},\tau\boldsymbol{I}_{k}\right),\quad\tau\sim\mathcal{IG}\left(\nu_{1},\nu_{2}\right),
\]
where $\nu_{1}$ and $\nu_{2}$ are fixed hyperparameters and $\mathcal{IG}\left(a,b\right)$
denotes an inverse gamma distribution with shape parameter $a$ and
rate parameter $b$.%
{} The third specification, referred to as the NIG-hetero prior, is
a normal-inverse-gamma prior with non-common element-specific hyperparameters:
\[
\theta_{j}|\tau_{j}\sim\mathcal{N}\left(0,\tau_{j}\right),\quad\tau_{j}\sim\mathcal{IG}\left(\nu_{1},\nu_{2}\right).
\]
For both the NIG-homo and NIG-hetero priors, the hyperparameters were
set to $\nu_{1}=2$ and $\nu_{2}=1$, rendering the priors moderately
informative. The hyperparameters were updated via a Gibbs sampling
step.

We considered combinations of different sample sizes and numbers of
covariates: $n\in\left\{ 100,1000\right\} $, $k\in\left\{ 5,20\right\} $.
For each experiment, a total of 200,000 draws were generated and the
last 100,000 draws were retained for analysis. Performance was evaluated
based on the multivariate effective sample sizes (mESS) \cite{Vats2019}.
Specifically, we computed the median values of mESS per iteration
(mESS/iter) and mESS per second (mESS/s) and root mean squared error
(RMSE) across 500 independent runs.\footnote{All the programs were executed in Matlab (R2025b) on an Ubuntu desktop
(24.04.3 LTS) running on an AMD Ryzen Threadripper 9980X (3.2 GHz).}

Table II summarizes the results for the Normal prior. Both DA--MCMC--Exact
and DA--MCMC--Approx substantially outperformed the benchmark algorithms
in terms of mESS per iteration and mESS per second across all experimental
settings. The performance gap widened as the dimension $k$ increased,
indicating that the proposed methods scale more effectively in higher-dimensional
problems. Between the two implementations, DA--MCMC--Exact achieved
higher mESS per iteration, reflecting its closer alignment with the
true quasi-posterior distribution. In addition, DA--MCMC--Exact
consistently exhibited superior computational efficiency, as indicated
by the higher mESS per second. The differences between the two implementations
were more pronounced for smaller samples ($n=100$) and larger dimensions
($k=20$), where the Exact version’s computational burden became more
evident. Overall, DA--MCMC--Exact outperformed DA--MCMC--Approx
on both the performance measures when the Normal prior was used. 

\begin{table}
\caption{Results for synthetic data (1) Normal prior}

\medskip{}

\centering{}%
\begin{tabular}{rrrrrr}
\hline 
$n$ & $k$ & MCMC & DA--MCMC & \multicolumn{2}{c}{Mod. DA--MCMC}\tabularnewline
\cline{5-6}
 &  &  &  & Exact & Approx\tabularnewline
\hline 
\multicolumn{6}{l}{(a) mESS/iter}\tabularnewline
\hline 
\multirow{2}{*}{100} & 5 & 0.040 & 0.060 & 0.848 & 0.372\tabularnewline
 & 20 & 0.004 & 0.010 & 0.421 & 0.061\tabularnewline
\hline 
\multirow{2}{*}{1,000} & 5 & 0.032 & 0.059 & 0.987 & 0.728\tabularnewline
 & 20 & 0.015 & 0.010 & 0.953 & 0.600\tabularnewline
\hline 
\multicolumn{6}{l}{(b) mESS/s}\tabularnewline
\hline 
\multirow{2}{*}{100} & 5 & 10,260 & 20,649 & 133,279 & 71,221\tabularnewline
 & 20 & 283 & 1375 & 20,924 & 7,371\tabularnewline
\hline 
\multirow{2}{*}{1,000} & 5 & 3,353 & 8,736 & 79,859 & 63,015\tabularnewline
 & 20 & 274 & 318 & 14,221 & 10,367\tabularnewline
\hline 
\multicolumn{6}{l}{(c) RMSE}\tabularnewline
\hline 
\multirow{2}{*}{100} & 5 & 0.180 & 0.183 & 0.182 & 0.192\tabularnewline
 & 20 & 0.811 & 0.340 & 0.338 & 0.367\tabularnewline
\hline 
\multirow{2}{*}{1,000} & 5 & 0.068 & 0.065 & 0.064 & 0.067\tabularnewline
 & 20 & 0.605 & 0.198 & 0.361 & 0.238\tabularnewline
\hline 
\end{tabular}
\end{table}

Table III presents the results for the NIG-homo prior. Compared with
that of the Normal prior case, the overall sampling efficiency of
the NIG-homo prior declined slightly, reflecting the additional uncertainty
introduced by the hyperparameter $\tau$. Nonetheless, both DA--MCMC--Exact
and DA--MCMC--Approx continued to outperform the benchmark algorithms
by substantial margins across all settings. For small samples ($n=100$)
and low dimensionality ($k=5$), both implementations achieved multivariate
effective sample sizes per iteration (mESS/iter) several times higher
than those of the baseline methods. As dimensionality increased ($k=20$),
efficiency gains diminished, and mESS values decreased noticeably,
highlighting the growing challenge of accurate sampling in higher-dimensional
parameter spaces under the hierarchical prior structure. In terms
of mESS per second, DA--MCMC--Approx again demonstrated superior
computational efficiency, particularly for $k=20$, where it outperformed
DA--MCMC--Exact. These results suggest that, although the hierarchical
shrinkage introduced by the NIG-homo prior increases computational
complexity, the proposed DA--MCMC framework remains effective and
stable across a wide range of sample sizes and model dimensions.

\begin{table}
\caption{Results for synthetic data (2) NIG-homo prior}

\medskip{}

\centering{}%
\begin{tabular}{rrrrrr}
\hline 
$n$ & $k$ & MCMC & DA--MCMC & \multicolumn{2}{c}{Mod. DA--MCMC}\tabularnewline
\cline{5-6}
 &  &  &  & Exact & Approx\tabularnewline
\hline 
\multicolumn{6}{l}{(a) mESS/iter}\tabularnewline
\hline 
\multirow{2}{*}{100} & 5 & 0.040 & 0.059 & 0.382 & 0.171\tabularnewline
 & 20 & 0.004 & 0.010 & 0.006 & 0.005\tabularnewline
\hline 
\multirow{2}{*}{1,000} & 5 & 0.031 & 0.059 & 0.686 & 0.555\tabularnewline
 & 20 & 0.015 & 0.010 & 0.061 & 0.051\tabularnewline
\hline 
\multicolumn{6}{l}{(b) mESS/s}\tabularnewline
\hline 
\multirow{2}{*}{100} & 5 & 5,474 & 9,274 & 41,110 & 21,408\tabularnewline
 & 20 & 217 & 891 & 428 & 733\tabularnewline
\hline 
\multirow{2}{*}{1,000} & 5 & 2,369 & 5,730 & 44,398 & 37,650\tabularnewline
 & 20 & 243 & 264 & 1,379 & 1,398\tabularnewline
\hline 
\multicolumn{6}{l}{(c) RMSE}\tabularnewline
\hline 
\multirow{2}{*}{100} & 5 & 0.202 & 0.203 & 0.214 & 0.229\tabularnewline
 & 20 & 0.742 & 0.399 & 0.470 & 0.482\tabularnewline
\hline 
\multirow{2}{*}{1,000} & 5 & 0.066 & 0.062 & 0.063 & 0.065\tabularnewline
 & 20 & 0.635 & 0.182 & 0.469 & 0.334\tabularnewline
\hline 
\end{tabular}
\end{table}

The results for the NIG-hetero prior are summarized in Table IV. Consistent
with the NIG-homo case, overall efficiency decreased relative to the
Normal prior, reflecting the additional complexity of sampling when
each coefficient is assigned an individual variance parameter. Nonetheless,
both DA--MCMC--Exact and DA--MCMC--Approx continued to substantially
outperform the benchmark methods across all scenarios. In terms of
mESS per iteration, DA--MCMC--Exact tended to yield slightly higher
values, particularly in lower-dimensional settings ($k=5$), indicating
that the richer hierarchical structure did not prevent effective exploration
of the posterior distribution. However, in higher dimensions ($k=20$),
the efficiency gap between DA--MCMC--Exact and DA--MCMC--Approx
narrowed, with the latter showing a modest advantage in mESS per second
due to its reduced computational burden. Overall, DA--MCMC--Approx
achieved a favorable balance between efficiency and stability, even
under the more flexible, heterogeneous prior structure. These results
confirm that the proposed framework remains robust when extended to
priors imposing coefficient-specific shrinkage, such as those used
in high-dimensional regression and sparse modeling contexts.
\begin{table}
\caption{Results for synthetic data (3) NIG-hetero prior}

\medskip{}

\centering{}%
\begin{tabular}{rrrrrr}
\hline 
$n$ & $k$ & MCMC & DA--MCMC & \multicolumn{2}{c}{Mod. DA--MCMC}\tabularnewline
\cline{5-6}
 &  &  &  & Exact & Approx\tabularnewline
\hline 
\multicolumn{6}{l}{(a) mESS/iter}\tabularnewline
\hline 
\multirow{2}{*}{100} & 5 & 0.041 & 0.059 & 0.361 & 0.212\tabularnewline
 & 20 & 0.004 & 0.010 & 0.085 & 0.019\tabularnewline
\hline 
\multirow{2}{*}{1,000} & 5 & 0.031 & 0.059 & 0.664 & 0.565\tabularnewline
 & 20 & 0.014 & 0.010 & 0.493 & 0.377\tabularnewline
\hline 
\multicolumn{6}{l}{(b) mESS/s}\tabularnewline
\hline 
\multirow{2}{*}{100} & 5 & 4,914 & 8,020 & 35,083 & 23,373\tabularnewline
 & 20 & 188 & 740 & 4,033 & 1,605\tabularnewline
\hline 
\multirow{2}{*}{1,000} & 5 & 2,221 & 5,216 & 39,128 & 36,035\tabularnewline
 & 20 & 221 & 244 & 7,236 & 6,098\tabularnewline
\hline 
\multicolumn{6}{l}{(c) RMSE}\tabularnewline
\hline 
\multirow{2}{*}{100} & 5 & 0.197 & 0.196 & 0.206 & 0.220\tabularnewline
 & 20 & 0.710 & 0.358 & 0.377 & 0.459\tabularnewline
\hline 
\multirow{2}{*}{1,000} & 5 & 0.058 & 0.059 & 0.060 & 0.062\tabularnewline
 & 20 & 0.616 & 0.173 & 0.499 & 0.315\tabularnewline
\hline 
\end{tabular}
\end{table}

Panels (c) of Tables II, III, and IV present the RMSEs. A similar
pattern emerges across all three tables: in the low-dimensional cases
($k=5$), the RMSEs of the four methods were comparable, whereas in
the higher-dimensional cases ($k=20$), the RMSEs of the standard
MCMC were larger than those of the DA--MCMC--type algorithms. This
suggests that the DA--MCMC--type algorithms facilitate more stable
inference. Although the DA--MCMC produced smaller RMSEs in the more
challenging cases, this does not necessarily indicate good performance,
as the small mESS/iter and mESS/s values suggest. In these settings,
the DA--MCMC tended to become trapped in localized regions of the
parameter space, often near the posterior modes.

Across all prior specifications and experimental settings, the proposed
DA--MCMC algorithms consistently outperformed the benchmark methods
in both sampling efficiency and computational speed. The DA--MCMC--Exact
variant achieved the highest per-iteration efficiency, whereas DA--MCMC--Approx
offered superior overall performance measured by effective sample
size per second. The relative advantage of the Approx version became
more pronounced as dimensionality increased or sample size grew, underscoring
its scalability and numerical stability. Taken together, these results
demonstrate that the proposed framework provides a flexible and computationally
efficient tool for quasi-Bayesian inference across a wide range of
model and prior configurations.

In practice, the choice between the Exact and Approximate variants
depends on the computational cost of evaluating the moment function.
The Exact version is preferable when full quasi-likelihood evaluations
are relatively computationally inexpensive, as it maintains the precise
acceptance rule and typically mixes well. The Approximate version
becomes attractive when evaluations are costly or high-dimensional,
as its surrogate-based first stage can reduce computation substantially.
As a rough guideline, the Exact version suits low-cost settings, whereas
the Approx version is more efficient when full evaluations are the
primary bottleneck.

\section{Application to Real Data}

We applied the proposed approach to infer an IV regression model (\ref{eq:=000020IV=000020regression}).
The inference procedure followed the same framework as that used for
the linear regression model, with the only distinction being the transformation
of the quasi-likelihood. Specifically, the exponential term in the
quasi-likelihood was modified as follows:
\[
\exp\left(-\frac{n}{2}\bar{\boldsymbol{m}}\left(\boldsymbol{\theta}\right)^{\top}\boldsymbol{W}\left(\boldsymbol{\theta}\right)\bar{\boldsymbol{m}}\left(\boldsymbol{\theta}\right)\right)
\]
\[
=\exp\left(-\frac{n}{2}\left[\frac{1}{n}\boldsymbol{Z}^{\top}\left(\boldsymbol{y}-\boldsymbol{X}\boldsymbol{\theta}\right)\right]^{\top}\boldsymbol{W}\left[\frac{1}{n}\boldsymbol{Z}^{\top}\left(\boldsymbol{y}-\boldsymbol{X}\boldsymbol{\theta}\right)\right]\right)
\]
\[
\propto\exp\left(-\frac{n}{2}\left(\boldsymbol{\theta}-\hat{\boldsymbol{\theta}}^{\dagger}\right)^{\top}\boldsymbol{G}^{\top}\boldsymbol{W}\boldsymbol{G}\left(\boldsymbol{\theta}-\hat{\boldsymbol{\theta}}^{\dagger}\right)\right),
\]
where $\hat{\boldsymbol{\theta}}^{\dagger}=\left(\boldsymbol{Z}^{\top}\boldsymbol{X}\right)^{-1}\boldsymbol{Z}^{\top}\boldsymbol{y}$
and $\boldsymbol{G}=n^{-1}\boldsymbol{Z}^{\top}\boldsymbol{X}$. Notably,
when the IV regression is exactly identified, $\hat{\boldsymbol{\theta}}^{\dagger}$
coincides with the two-stage least squares estimator \cite{Greene2017}.
We employed the NIG-hetero prior with the same hyperparameters as
in Section III. 

We applied the IV regression to two real datasets. The first dataset,
denoted as AJR, was originally compiled by \cite{Acemoglu2001,Acemoglu2012}\footnote{https://www.openicpsr.org/openicpsr/project/112564/version/V1/view.}.
They investigated the effect of the risk of expropriation on gross
domestic product per capita. To address potential endogeneity in this
relationship, European settler mortality was used as an instrumental
variable. The specification also includes several control variables:
a constant term, the latitude of each country (and its square), and
dummy variables indicating whether the country is located in Africa
or Asia, as well as whether it belongs to the group of former British
colonies (Australia, Canada, New Zealand, and the United States).
The sample consists of $n=64$ observations, and the number of moment
conditions is $k=10$. 

The second dataset, denoted as Movies, originates from \cite{Dahl2009}.
We used the version provided in Chapter 12 of \cite{Stock2020}\footnote{https://www.princeton.edu/\textasciitilde mwatson/Stock-Watson\_4E/Stock-Watson-Resources-4e.html.}.
The data cover 516 weekends between 1995 and 2004 and record the number
of assaults across selected U.S. counties, along with national attendance
figures for highly violent films. We estimated the relationship between
weekend assault counts and film attendance using an instrumental variables
approach, where predicted attendance serves as an instrument for observed
attendance. The specification also includes a comprehensive set of
control variables, such as fixed effects for year and month, indicators
for holiday weekends, and multiple weather-related covariates. The
dimension of the inferential problem is characterized by $n=516$
and $k=36$.

The results are summarized in Table V. In both empirical applications,
the proposed DA--MCMC methods outperformed the benchmark algorithms
in sampling efficiency. For the AJR dataset, DA--MCMC--Exact achieved
the highest mESS per iteration (mESS/iter) and per second (mESS/s),
reflecting strong computational efficiency in a moderately sized,
exactly identified model. DA--MCMC--Approx also performed competitively,
providing substantial improvement over the baseline methods at a lower
computational cost. When DA--MCMC--Approx was used, the posterior
mean of the coefficient on expropriation risk was 1.09 with a corresponding
standard deviation of 0.20. This result is close to those reported
in Table 4 of \cite{Acemoglu2001}, confirming their findings using
moment-based quasi-Bayesian inference with a shrinkage prior.

\begin{table}
\caption{Results for real data}

\medskip{}

\centering{}%
\begin{tabular}{lrrrrrr}
\hline 
Data & $n$ & $k$ & MCMC & DA--MCMC & \multicolumn{2}{c}{Mod. DA--MCMC}\tabularnewline
\cline{6-7}
 &  &  &  &  & Exact & Approx\tabularnewline
\hline 
\multicolumn{7}{l}{(a) mESS/iter}\tabularnewline
\hline 
AJR & 64 & 10 & 0.004 & 0.007 & 0.015 & 0.004\tabularnewline
Movies & 516 & 36 & 0.004 & 0.004 & -- & 0.093\tabularnewline
\hline 
\multicolumn{7}{l}{(b) mESS/s}\tabularnewline
\hline 
AJR & 64 & 10 & 311 & 882 & 1,658 & 636\tabularnewline
Movies & 516 & 36 & 53 & 93 & -- & 5,449\tabularnewline
\hline 
\end{tabular}
\end{table}

For the Movies dataset, the advantage of DA--MCMC--Approx became
especially pronounced. The Exact version was computationally infeasible
in this higher-dimensional setting ($k=36$), whereas DA--MCMC--Approx
achieved exceptionally high efficiency, yielding an mESS/s more than
an order of magnitude greater than that of the benchmark algorithms.
These results demonstrate the scalability and robustness of the Approx
implementation in complex, high-dimensional empirical problems.

Overall, the empirical analyses reinforce the findings from the synthetic
data experiments. Both implementations of the proposed DA--MCMC framework
yielded substantial improvements in sampling efficiency compared to
conventional MCMC and DA--MCMC methods, even in realistic econometric
settings. The DA--MCMC--Exact variant provided the most precise
inference in low- to moderate-dimensional models, whereas the DA--MCMC--Approx
variant proved considerably more scalable and computationally stable
in higher-dimensional applications. These results highlight the practical
versatility of the DA--MCMC framework for quasi-Bayesian inference
in diverse empirical contexts.

\section{Discussion}

This paper introduces a computationally efficient framework for quasi-Bayesian
inference based on the DA--MCMC algorithm. By leveraging the linear
structure of moment conditions, the method constructs proposal distributions
that closely approximate the conditional posterior, improving both
mixing and computational performance. Two implementations---DA--MCMC--Exact
and DA--MCMC--Approx---were developed to balance computational
efficiency and numerical stability. Simulation studies using synthetic
data demonstrated substantial gains in sampling efficiency compared
to standard MCMC and conventional DA--MCMC methods, while empirical
applications to real datasets confirmed the scalability and robustness
of the approach in practical settings.

Future research could extend this framework to nonlinear and overidentified
models, refine the surrogate kernel, and explore integration with
modern variational or sequential inference techniques. Overall, the
proposed DA--MCMC framework provides a versatile and computationally
tractable tool for quasi-Bayesian analysis in complex statistical
models.

\appendix{}

The main term inside the exponential function in the second line of
(\ref{eq:=000020transformed=000020quasi-posterior}), after removing
the term $-2/n$, can be written as follows:

\[
\left[\frac{1}{n}\boldsymbol{X}^{\top}\left(\boldsymbol{y}-\boldsymbol{X}\boldsymbol{\theta}\right)\right]^{\top}\boldsymbol{W}\left[\frac{1}{n}\boldsymbol{X}^{\top}\left(\boldsymbol{y}-\boldsymbol{X}\boldsymbol{\theta}\right)\right]
\]
\[
=C-2\left(\frac{1}{n}\boldsymbol{X}^{\top}\boldsymbol{X}\boldsymbol{\theta}\right)^{\top}\boldsymbol{W}\left(\frac{1}{n}\boldsymbol{X}^{\top}\boldsymbol{y}\right)
\]
\[
\qquad+\left(\frac{1}{n}\boldsymbol{X}^{\top}\boldsymbol{X}\boldsymbol{\theta}\right)^{\top}\boldsymbol{W}\left(\frac{1}{n}\boldsymbol{X}^{\top}\boldsymbol{X}\boldsymbol{\theta}\right)
\]
\[
=C-2\boldsymbol{\theta}^{\top}\left(\frac{1}{n}\boldsymbol{X}^{\top}\boldsymbol{X}\right)^{\top}\boldsymbol{W}
\]
\[
\quad\times\left(\frac{1}{n}\boldsymbol{X}^{\top}\boldsymbol{X}\boldsymbol{\theta}\right)\left(\frac{1}{n}\boldsymbol{X}^{\top}\boldsymbol{X}\boldsymbol{\theta}\right)^{-1}\left(\frac{1}{n}\boldsymbol{X}^{\top}\boldsymbol{y}\right)
\]
\[
\qquad+\boldsymbol{\theta}^{\top}\left(\frac{1}{n}\boldsymbol{X}^{\top}\boldsymbol{X}\right)^{\top}\boldsymbol{W}\left(\frac{1}{n}\boldsymbol{X}^{\top}\boldsymbol{X}\right)\boldsymbol{\theta}
\]
\[
=\left(C-\left(\boldsymbol{\theta}^{\dagger}\right)^{\top}\boldsymbol{G}^{\top}\boldsymbol{W}\boldsymbol{G}\boldsymbol{\theta}^{\dagger}\right)+\left(\boldsymbol{\theta}^{\dagger}\right)^{\top}\boldsymbol{G}^{\top}\boldsymbol{W}\boldsymbol{G}\boldsymbol{\theta}^{\dagger}
\]
\[
\qquad-2\boldsymbol{\theta}^{\top}\boldsymbol{G}^{\top}\boldsymbol{W}\boldsymbol{G}\boldsymbol{\theta}^{\dagger}+\boldsymbol{\theta}^{\top}\boldsymbol{G}^{\top}\boldsymbol{W}\boldsymbol{G}\boldsymbol{\theta},
\]
where $C=\left(n^{-1}\boldsymbol{X}^{\top}\boldsymbol{y}\right)^{\top}\boldsymbol{W}\left(n^{-1}\boldsymbol{X}^{\top}\boldsymbol{y}\right)$.

\bibliographystyle{ieeetr}
\bibliography{reference}

\end{document}